\documentstyle[12pt]{article}

\begin{document}

\title{Non-commutative time-frequency tomography}
\author{{V.I. Man'ko\thanks{%
on leave from the P. N. Lebedev Physical Institute, Moscow, Russia} and
R.Vilela Mendes} \\
{\it Grupo de F\'\i sica--Matem\'atica, Complexo Interdisciplinar}\\
{\it Universidade de Lisboa}\\
{\it Av. Prof. Gama Pinto, 2, 1699 Lisboa Codex, Portugal}}
\date{}
\maketitle

\begin{abstract}
The characterization of non-stationary signals requires joint time and
frequency information. However, time $(t)$ and frequency $(\omega )$ being
non-commuting variables, there cannot be a joint probability density in the $%
(t,\omega )$ plane and the time-frequency distributions, that have been
proposed, have difficult interpretation problems arising from negative or
complex values and spurious components. As an alternative we propose to
obtain time-frequency information by looking at the marginal distributions
along rotated directions in the $(t,\omega )$ plane. The rigorous
probability interpretation of the marginal distributions avoids all
interpretation ambiguities. Applications to signal analysis and signal
detection are discussed as well as an extension of the method to other pairs
of non-commuting variables.
\end{abstract}

\section{Introduction}

Non-stationary signals have a time-dependent spectral content, therefore, an
adequate characterization of these signals requires joint time and frequency
information. Among the many time-frequency (quasi)distributions\cite{Cohen1} 
\cite{Faye} that have been proposed, {\it Wigner-Ville}'s (WV)\cite{Wigner} 
\cite{Ville} 
\begin{equation}
\label{1.1}W\left( t,\omega \right) =\int f\left( t+\frac u2\right)
f^{*}\left( t-\frac u2\right) e^{-i\omega u}\,du\, 
\end{equation}
for an analytic signal $f\left( t\right) $, is considered to be optimal in
the sense that it satisfies the marginals, it is time-frequency shift
invariant and it possesses the least amount of spread in the time-frequency
plane.

However, the WV distribution has, in general, positive and negative values
and may be non-zero in regions of the time-frequency plane where either the
signal or its Fourier transform vanish. Therefore, despite the fact that the
WV distribution is an accurate mathematical characterization of the signal,
in the sense that it can be inverted by 
\begin{equation}
\label{1.2}f(t)f^{*}(t^{^{\prime }})=\frac 1{2\pi }\int W\left( \frac{%
t+t^{^{\prime }}}2,\omega \right) e^{i\omega (t-t^{^{\prime }})}\,d\omega \, 
\end{equation}
its interpretation for signal detection and recognition is no easy matter,
because of the negative and ''spurious'' components. The origin of this
problem lies in the fact that $t$ and $\omega $ being non-commuting
variables, they cannot be simultaneously specified with absolute accuracy
and, as a result, there cannot be a joint probability density in the
time-frequency plane. Therefore no joint distribution, even if positive\cite
{Cohen2}, may be interpreted as a probability density.

Looking back at the original motivation leading to the construction of the
time-frequency distributions, namely the characterization of non-stationary
signals, we notice that we are asking for more than we really need. To
characterize a non-stationary signal what we need is time and
frequency-dependent information, not necessarily a joint probability
density, a mathematical impossibility for non-commuting variables. The
solution is very simple. The time density $\left| f(t)\right| ^2$ projects
the signal intensity on the time axis and the spectral density $\left|
f(\omega )\right| ^2$ projects on the frequency axis. To obtain the required
time-frequency information, all we need is a family of time and frequency
functions $s_\xi (t,\omega )$, depending on a parameter $\xi $, which
interpolates between time and frequency. Projecting the signal intensity on
this variable, that is, computing the density along the $s_\xi -$axis, one
obtains a function 
\begin{equation}
\label{1.3}M(s,\xi )=\left| f(s_\xi )\right| ^2 
\end{equation}
that has, for each $\xi $, a probability interpretation. The simplest choice
for $s_\xi $ is a linear combination 
\begin{equation}
\label{1.4}s=\mu t+\nu \omega 
\end{equation}
the parameter $\xi $ being the pair $(\mu ,\nu )$. For definiteness we may
choose 
\begin{equation}
\label{1.5}
\begin{array}{c}
\mu = 
\frac{\cos \theta }T \\ \nu =\frac{\sin \theta }\Omega 
\end{array}
\end{equation}
$T,\Omega $ being a reference time and a reference frequency adapted to the
signal to be studied. The function $M(s,\theta )$ interpolates between $%
\left| f(t)\right| ^2$ and $\left| f(\omega )\right| ^2$ and, as we will
prove below, contains a complete description of the signal. For each $\theta 
$ the function $M(s,\theta )$ is strictly positive and being a bona-fide
probability (in $s$) causes no interpretation ambiguities. A similar
approach has already been suggested for quantum optics\cite{Vogel} and
quantum mechanics\cite{Mancini1} \cite{Mancini2} \cite{Mancini3}, the
non-commuting variable pairs being respectively the quadrature phases $%
(a_r,a_i)$ and the position-momentum $(q,p)$.

This approach, in which to reconstruct an object, be it a signal in signal
processing or a wave function in quantum mechanics, one looks at its
probability projections on a family of rotated axis, is similar to the {\it %
computerized axial tomography} (CAT) method. The basic difference is that in
CAT scans one deals with a pair $(x,y)$ of commuting position variables and
here we deal with a plane defined by a pair of non-commuting variables. For
this reason we call the present approach {\it non-commutative tomography}
(NCT).

The paper is organized as follows. In Section 2 we construct the NCT\ signal
transform and show its positivity and normalization properties. We also
establish the invertibility of the transformation, which shows that it
contains a complete description of the signal and establish its relation to
the WV distribution. Because the NCT transform involves the square of the
absolute value of a linear functional of the signal, it is actually easier
to compute than bilinear transforms like WV.

In Section 3 we work out the analytical form of the NCT transform for some
signals and also display the $M(s,\theta )$ in some examples. We also 
deal with the problem of using NCT to detect the presence of signals in
noise for small {\it signal to noise ratios} (SNR). Here the essential
observation is that, for small SNR, the signal may be difficult to detect
along $t$ or $\omega $, however, it is probable that there are other
directions on the $(t,\omega )$ plane along which detection might be easier.
It is the consistent occurrence of many such directions that supplies the
detection signature.

Finally in Section 4 we point out that the NCT approach may also be used for
other pairs of non-commuting variables of importance in signal processing.
As an example we work out the relevant formulas for the scale-frequency pair.

\section{Non-commutative time-frequency tomography}

Because the Fourier transform of a characteristic function is a probability
density, we compute the marginal distribution for the variable $s=\mu t+\nu
\omega $ using the characteristic function method. Frequency and time are
operators acting in the Hilbert space of analytic signals and, in the
time-representation, the frequency operator is $\omega =-i\partial /\partial
t$~. The characteristic function $C(k)$ is 
\begin{equation}
\label{2.1}C(k)=\langle e^{ik\left( \mu t+\nu \omega \right) }\rangle =\int
f^{*}(t)\,e^{ik\left( \mu t-i\nu \partial /\partial t\right) }\,f(t)\,dt\, 
\end{equation}
where $f(t)$ is a normalized signal%
$$
\int \left| f(t)\right| ^2dt=1 
$$
The Fourier transform of the characteristic function is a probability
density 
\begin{equation}
\label{2.2}M\left( s,\mu ,\nu \right) =\frac 1{2\,\pi }\int
C(k)e^{-iks}\,dk\, 
\end{equation}
After some algebra, one obtains the marginal distribution~(\ref{2.2}) in
terms of the analytical signal 
\begin{equation}
\label{2.3}M\left( s,\mu ,\nu \right) =\frac 1{2\,\pi |\nu |}\left| \int
\exp \left[ \frac{i\mu t^2}{2\,\nu }-\frac{its}\nu \right] f(t)\,dt\right|
^2 
\end{equation}
with normalization 
\begin{equation}
\label{2.4}\int M\left( s,\mu ,\nu \right) \,ds=1\, 
\end{equation}
For the case $\mu =1,\,\nu =0,$ it gives the distribution of the analytic
signal in the time domain 
\begin{equation}
\label{2.5}M\left( t,1,0\right) =|f(t)|^2 
\end{equation}
and for the case $\mu =0,\,\nu =1,$ the distribution of the analytic signal
in the frequency domain 
\begin{equation}
\label{2.6}M\left( \omega ,0,1\right) =|f(\omega )|^2 
\end{equation}

The family of marginal distributions $M(s,\mu ,\nu )$ contains complete
information on the analytical signal. This may be shown directly. However it
is more interesting to point out that there is an invertible transformation
connecting $M(s,\mu ,\nu )$ to the Wigner-Ville quasidistribution, namely 
\begin{equation}
\label{2.7}M\left( s,\mu ,\nu \right) =\int \exp \left[ -ik(s-\mu t-\nu
\omega )\right] W(t,\omega )\,\frac{dk\,d\omega \,dt}{(2\pi )^2} 
\end{equation}
and 
\begin{equation}
\label{2.8}W(t,\omega )=\frac 1{2\pi }\int M\left( s,\,\mu ,\,\nu \right)
\exp \left[ -i\left( \mu t+\nu \omega -s\right) \right] \,d\mu \,d\nu \,ds 
\end{equation}
Therefore, because the WV quasidistribution has complete information, in the
sense of Eq.(\ref{1.2}), so has $M(s,\mu ,\nu )$.

\section{Examples}

We compute the NCT transform $M(s,\mu ,\nu )$ for some analytic signals:

(i){\it \ A complex Gaussian signal} 
\begin{equation}
\label{3.1}f(t)=\left( \frac \alpha \pi \right) ^{1/4}\exp \left[ -\frac
\alpha 2\,t^2+i\,\frac \beta 2\,t^2+i\,\omega _0t\right] 
\end{equation}
It has the properties 
\begin{equation}
\label{3.2}\langle t\rangle =0,\qquad \langle \omega \rangle =\omega _0\, 
\end{equation}
\begin{equation}
\label{3.3}
\begin{array}{ccccc}
\sigma _t^2 & = & \langle t^2\rangle -\langle t\rangle ^2 & = & \frac
1{2\alpha } \\ 
\sigma _\omega ^2 & = & \langle \omega ^2\rangle -\langle \omega \rangle ^2
& = & \frac{\alpha ^2+\beta ^2}{2\alpha } \\ r & = & \frac{2^{-1}\langle
t\omega +\omega t\rangle -\langle t\rangle \langle \omega \rangle }{\sigma
_\omega \sigma _t} & = & \frac \beta {\sqrt{\alpha ^2+\beta ^2}} 
\end{array}
\end{equation}
This signal minimizes the Robertson--Schr\"odinger uncertainty relation 
\begin{equation}
\label{3.4}\sigma _\omega ^2\sigma _t^2\geq \frac 14\,\frac 1{1-r^2}\, 
\end{equation}
In quantum mechanics, it corresponds to a correlated coherent state~\cite
{Kurm80} \cite{Bham}.

The NCT transform is 
\begin{equation}
\label{3.5}M\left( s,\,\mu ,\,\nu \right) =\frac 1{\sqrt{2\,\pi \sigma _s^2}%
}\,\exp \left[ -\frac{\left( s-\overline{s}\right) ^2}{2\,\sigma _s^2}%
\right] 
\end{equation}
with parameters 
\begin{equation}
\label{3.6}
\begin{array}{ccl}
\sigma _s^2 & = & \frac 1{2\alpha }\left| \nu \left( \alpha -i\beta \right)
-i\mu \right| ^2 \\ 
\overline{s} & = & \omega _0\nu 
\end{array}
\end{equation}
For the case of $\mu =\frac{\cos \,\theta }T,\,\nu =\frac{\sin \,\theta }%
\Omega ,$ Eq.(\ref{3.6}) shows how the initial Gaussian evolves along the $%
\theta $ axis, changing its maximum and width 
\begin{equation}
\label{3.7}
\begin{array}{ccl}
\sigma _s^2 & = & \frac 1{2\alpha }\left| 
\frac{\sin \,\theta }\Omega \left( \alpha -i\beta \right) -i\frac{\cos
\,\theta }T\right| ^2 \\ \overline{s} & = & \omega _0\frac{\sin \,\theta }%
\Omega 
\end{array}
\end{equation}
Thus, we have squeezing in the quadrature components and their correlation.
In the case $\beta =0,$ one has a purely squeezed state\cite{Yuen76} \cite
{Walls}, which minimizes the Heisenberg uncertainty relation 
\begin{equation}
\label{3.8}\sigma _\omega ^2\sigma _t^2\geq \frac 14\, 
\end{equation}

(ii) {\it A normalized superposition of two Gaussian signals} 
\begin{equation}
\label{3.9}f(t)=N_s\left\{ A_1f_1(t)+A_2f_2(t)\right\} 
\end{equation}
where $f_i(t)$ is 
\begin{equation}
\label{3.10}f_i(t)=N_i\exp \left[ -a_it^2+b_it\right] ,\qquad i=1,2\, 
\end{equation}
and 
\begin{equation}
\label{3.11}N_i=\left[ \frac{a_i+a_i^{*}}\pi \right] ^{1/4}\exp \left[
-\frac 18\,\frac{\left( b_i+b_i^{*}\right) ^2}{a_i+a_i^{*}}\right] 
\end{equation}
The superposition coefficients being complex numbers, the normalization
constant $N_s$ reads 
\begin{equation}
\label{3.12}N_s=\left( |A_1|^2+|A_2|^2+2\,\textnormal{Re}\left[ A_1A_2^{*}\int
f_1(t)\,f_2^{*}(t)\,dt\right] \right) ^{-1/2} 
\end{equation}
Computing the marginal distribution $M\left( s,\,\mu ,\,\nu \right) $ by Eq.(%
\ref{2.3}) we arrive at a combination of three Gaussian terms 
\begin{equation}
\label{3.13}
\begin{array}{ccl}
M\left( s,\,\mu ,\,\nu \right) & = & N_s^2\{|A_1|^2M_1\left( s,\,\mu ,\,\nu
\right) +|A_2|^2M_2\left( s,\,\mu ,\,\nu \right) \\  
&  & +2\,\textnormal{Re}[A_1A_2^{*}M_{12}\left( s,\,\mu ,\,\nu \right) ]\} 
\end{array}
\end{equation}
where we have the contribution of two real Gaussian terms 
\begin{equation}
\label{3.14}M_i\left( s,\mu ,\nu \right) =\frac 1{\sqrt{2\pi \sigma _i^2}%
}\,\exp \left[ -\frac{\left( s-\overline{s}_i\right) ^2}{2\sigma _i^2}%
\right] ,\qquad i=1,2, 
\end{equation}
and the superposition of two complex Gaussians 
\begin{equation}
\label{3.15}M_{12}\left( s,\mu ,\nu \right) =\frac{n_{12}}{\sqrt{2\pi \sigma
_{12}^2}}\,\exp \left[ -\frac{\left( s-\overline{s}_{12}\right) ^2}{2\sigma
_{12}^2}\right] 
\end{equation}
The parameters of the real Gaussians are the dispersion 
\begin{equation}
\label{3.16}\sigma _i^2=2\,\frac{\left| \nu a_i-\frac{i\mu }2\right| ^2}{%
a_i+a_i^{*}} 
\end{equation}
and mean 
\begin{equation}
\label{3.17}\overline{s}_i=\nu \,\frac{\textnormal{Im}\left( b_ia_i^{*}\right) +%
\textnormal{Re}\left( \frac \mu {2\nu }b_i\right) }{\textnormal{Re }a_i}\, 
\end{equation}
The parameters of the complex Gaussian are 
\begin{equation}
\label{3.18}\sigma _{12}^2=2\nu ^2\,\frac{\left( a_1-\frac{i\mu }{2\nu }%
\right) \left( a_2^{*}+\frac{i\mu }{2\nu }\right) }{a_1+a_2^{*}} 
\end{equation}
and 
\begin{equation}
\label{3.19}\overline{s}_{12}=\frac{i\nu }{a_1+a_2^{*}}\left[ b_2^{*}\left(
a_1-\frac{i\mu }{2\nu }\right) -b_1\left( a_2^{*}+\frac{i\mu }{2\nu }\right)
\right] 
\end{equation}
and the complex amplitude $n_{12}$ of the complex Gaussian contribution is
\begin{equation}
\label{ss12}n_{12}=\frac{\sigma _{12}}{\sqrt{2\pi }|\nu |}\exp \left[ \frac
14\left( \frac{b_1^2}{a_1-\frac{i\mu }{2\nu }}+\frac{b_2^{*2}}{a_2^{*}+\frac{%
i\mu }{2\nu }}\right) +\frac{\overline{s}_{12}^2}{2\sigma _{12}^2}\right] 
\end{equation}

(iii) {\it Finite-time signals}

Here we consider signals 
\begin{equation}
\label{3.21}f_i(t)=N_ie^{-a_it^2+b_it}\,,\qquad t_{2i}\leq t\leq t_{1i}\, 
\end{equation}
which vanish for all other times and compute the NCT for one signal and for
the superposition of two such signals. The parameters $a_i$ and $b_i$ are
complex numbers. The normalization constant is 
\begin{eqnarray}
{\cal N}_i&=&\sqrt {a_i+a_i^*}\exp \left [-\frac {\left(b_i+b_i^*\right)^2}
{4\left(a_i+
a_i^*\right)}\right]\Big|\frac {\sqrt \pi}{2}\left[\mbox {erfc}\left(
\sqrt {a_i+a_i^*}\left [t_{2i}-\frac {b_i+b_i^*}{2\left(a_i+
a_i^*\right)}\right]\right)\right.\nonumber\\
&&-\left.\mbox {erfc}\left(
\sqrt {a_i+a_i^*}\left [t_{1i}-\frac {b_i+b_i^*}{2\left(a_i+
a_i^*\right)}\right]\right)\right]\Big|^{-1/2}\label{3.22}
\end{eqnarray}
where erfc is the function 
\begin{equation}
\label{3.23}\,\textnormal{erfc}\left( y\right) =\frac 2{\sqrt{\pi }}\int_y^\infty
e^{-x^2}\,dx\, 
\end{equation}
Using Eq.(\ref{2.3}), we arrive at the following marginal distribution 
\begin{eqnarray}
M_i\left(s,\mu,\nu\right)&=&\frac {{\cal N}_i^2}{8|\nu |}\Big|
\mbox {erfc}\left(
\sqrt {a_i-\frac {i\mu}{2\nu}}\left [t_{2i}-\frac {\nu b_i-is}{2\nu a_i
-i\mu}\right]\right)\nonumber\\
&&-\mbox {erfc}\left(\sqrt {a_i-\frac {i\mu}{2\nu}}
\left [t_{1i}-\frac {\nu b_i-is}{2\nu a_i-i\mu}
\right]\right)\Big|^2\label{3.24}
\end{eqnarray}
In the limit $t_{1i}\rightarrow -\infty ,~t_{2i}\rightarrow \infty ,$ the
marginal distributions~(\ref{3.24}) reduce to the Gaussian distribution given
by~(\ref{3.14}). In the case $a_i=0,\,b_i=i\omega _i,$ the distribution~(\ref
{3.24}) describes a sinusoidal signal of finite duration. The normalization
constant takes the limit value 
\begin{equation}
\label{3.25}N_i\Longrightarrow \left( t_{2i}-t_{1i}\right) ^{-1/2} 
\end{equation}
For a superposition of two finite-time signals%
$$
f(t)=N_s\left\{ A_1f_1(t)+A_2f_2(t)\right\} 
$$
with the signals $f_1(t)$ and $f_2(t)$ as in (\ref{3.21}), the normalization
constant is given by Eq.(\ref{3.12}) with overlap integral 
\begin{eqnarray}
\int_{t_a}^{t_b}f_1(t)\,f_2^*(t)\,dt&=&{\cal N}_1{\cal N}_2
\frac {\sqrt \pi}{2\sqrt {a_1+a_2^*}}\,\exp \left [\frac {\left(b_i
+b_i^*\right)^2}{4\left(a_i+a_i^*\right)}\right]\nonumber\\
&&\left\{\mbox {erfc}\left(\sqrt {a_1+a_2^*}\left [t_a-\frac {b_1+b_2^*}
{2\left(a_1+a_2^*\right)}\right]\right)\right.\nonumber\\
&&-\left.\mbox {erfc}\left(
\sqrt {a_1+a_2^*}\left [t_b-\frac {b_1+b_2^*}{2\left(a_1+
a_2^*\right)}\right]\right)\right\}
\label{3.26}
\end{eqnarray}

The marginal distribution for the superposition signal has the same form as
Eq.~(\ref{3.13}) but with a changed normalization constant, the
distributions $M_1\left( s,\mu ,\nu \right) $ and $M_2\left( s,\mu ,\nu
\right) $ given by Eq.~(\ref{3.24}), and an interference term $M_{12}\left(
s,\mu ,\nu \right) $ 
\begin{eqnarray}
M_{12}\left(s,\mu,\nu\right)&=&\frac {{\cal N}_1{\cal N}_2}{8|\nu |}
\left\{\mbox {erfc}\left(\sqrt {a_1-\frac {i\mu}{2\nu}}\left [t_{21}
-\frac {\nu b_1-is}{2\nu a_1-i\mu}\right]\right)\right.\nonumber\\
&&-\left.\mbox {erfc}\left(\sqrt {a_1-\frac {i\mu}{2\nu}}
\left [t_{11}-\frac {\nu b_1-is}{2\nu a_1-i\mu}
\right]\right)\right\}\nonumber\\
&&\times 
\left\{\mbox {erfc}\left(\sqrt {a_2-\frac {i\mu}{2\nu}}\left [t_{22}
-\frac {\nu b_2-is}{2\nu a_2-i\mu}\right]\right)\right.\nonumber\\
&&-\left.\mbox {erfc}\left(\sqrt {a_2-\frac {i\mu}{2\nu}}
\left [t_{12}-\frac {\nu b_2-is}{2\nu a_2-i\mu}
\right]\right)\right\}^*
\label{3.27}
\end{eqnarray}
The case $a_2=0$ corresponds to the combination of a finite time chirp and a
finite time sinusoidal signal shown in one of the figures below.

(iv) {\it Graphical illustrations}

We have plotted $M\left( s,\mu ,\nu \right) $ for some signals. In all cases
we use $\mu $ and $\nu $ as in Eq.(\ref{1.5}) with $T=1$ and $\Omega =10$.
All signals are finite time signals and in each case we display a
three-dimensional and a contour plot.

\# Figs 1a,b. The signal is 
\begin{equation}
\label{3.31}f(t)=\left\{ 
\begin{array}{lcc}
e^{-i20t}+e^{i10t} & \bigskip\  & t\in (0,1) \\ 
0 & \bigskip\  & t\notin (0,1) 
\end{array}
\right\} 
\end{equation}
Although the number of periods, during which is signal is on, is relatively
small, the two contributing frequencies are clearly seen in the separating
ridges.

\# Figs 2a,b. The signal is 
\begin{equation}
\label{3.32}f(t)=\left\{ 
\begin{array}{lcl}
e^{-i20t} & \bigskip\  & t\in (0,\frac 14) \\ 
0 & \bigskip\  & t\in (\frac 14,\frac 34) \\ 
e^{i10t} & \bigskip\  & t\in (\frac 34,1) 
\end{array}
\right\} 
\end{equation}
Once again the contributions separate as $\theta $ grows, but notice the
intermediate interference region which is a signature of the time-sequence
of the frequencies occurrence and of their relative phase.

\# Figs 3a,b. The signal is 
\begin{equation}
\label{3.33}f(t)=\left\{ 
\begin{array}{lcc}
e^{-i\left( 20t+10t^2\right) }+e^{i10t} & \bigskip\  & t\in (0,1) \\ 
0 & \bigskip\  & t\notin (0,1) 
\end{array}
\right\} 
\end{equation}
Contrasts the signature shapes of a chirp contribution and a regular
sinusoidal pulse.

Notice that all $M(s,\theta )$ values have a probability interpretation.
Therefore all peaks or oscillations have a direct physical meaning and, as
opposed to the time-frequency quasidistributions, we need not worry about
spurious effects. This is particularly important for the detection of
signals in noise, as we will see in the next example.

(v) {\it Detection of noisy signals by NCT}

In Fig.4a and 4b we have plotted a time signal $f(t)$ and its spectral
density $\left| f(\omega )\right| ^2$. It is really very hard to decide,
from these plots, where this signal might have originated from. Now we plot
the NCT transform (Fig.4c) and its contour plot (Fig.4d) with the
normalization $T=1$ and $\Omega =1000$. It still looks quite complex but, 
among all the peaks, one may clearly see a sequence of small peaks connecting 
a time around $0.5$ to a frequency around $200$.

In fact the signal was generated as a superposition of a normally
distributed random amplitude and random phase noise with a sinusoidal signal
of the same average amplitude but operating only during the time interval
$(0.45,0.55)$%
. This means that, during the observation time, the signal to noise power
ratio is $1/10$. The signature that the signal leaves on the NCT transform
is a manifestation of the fact that, despite its low SNR, there is a
number of particular directions in the $(t,\omega )$ plane along
which detection happens to be more favorable. The reader may convince
himself of the soundness of this interpretation by repeating the experiment
with different noise samples and noticing that each time the coherent peaks
appear at different locations, but the overall geometry of the ridge is the
same.

Of course, to rely on a ridge of small peaks for detection purposes only
makes sense because the rigorous probability interpretation of $M(s,\theta )$
renders the method immune to spurious effects.

\section{NCT for other non-commuting pairs. The time-scale and
frequency--scale cases}

The method may also be applied to other pairs of non-commuting variables for
which, as in the time-frequency case, there cannot be a joint probability
density. Consider the pair time-scale, where scale is the operator 
\begin{equation}
\label{4.1}D=\frac 12\left( t\omega +\omega t\right) =\omega t+\frac i2 
\end{equation}
In the plane $(t,D)$ we consider the linear combination 
\begin{equation}
\label{4.2}s_1=\mu t+\nu D=\frac{\cos \theta }Tt+\nu D 
\end{equation}
The relevant characteristic function is 
\begin{equation}
\label{4.3}
\begin{array}{c}
C_{\mu \nu }^{(1)}(k)=\left\langle e^{ik\left( \mu t+\nu D\right)
}\right\rangle =\int f^{*}(t)e^{ik\left( \mu t+\nu D\right) }f(t)dt \\ 
=\int f^{*}(e^{-\frac{k\nu }2}t)e^{i2\frac \mu \nu \sinh \left( \frac{k\nu }%
2\right) }f(e^{\frac{k\nu }2}t)dt 
\end{array}
\end{equation}
and the NCT transform is, as before, the Fourier transform of $C_{\mu \nu
}^{(1)}(k)$%
$$
M^{(1)}\left( s_1,\mu ,\nu \right) =\frac 1{2\,\pi }\int C_{\mu \nu
}^{(1)}(k)e^{-iks_1}\,dk\, 
$$
leading to 
\begin{equation}
\label{4.4}
\begin{array}{ccc}
M^{(1)}\left( s_1,\mu ,\nu \right) & = & \frac 1{2\,\pi \left| \nu \right|
}\left| \int_{t>0} 
\frac{dt}{\sqrt{t}}f(t)e^{i\left( \frac \mu \nu t-\frac{s_1}\nu \log
t\right) }\right| ^2+ \\  &  & \frac 1{2\,\pi \left| \nu \right| }\left|
\int_{t<0}\frac{dt}{\sqrt{\left| t\right| }}f(t)e^{i\left( \frac \mu \nu t-%
\frac{s_1}\nu \log \left| t\right| \right) }\right| ^2 
\end{array}
\end{equation}

For the pair frequency-scale, $(\omega ,D)$, we obtain similarly 
\begin{equation}
\label{4.5}s_2=\mu \omega +\nu D=\frac{\cos \theta }\Omega \omega +\nu D 
\end{equation}
\begin{equation}
\label{4.6}
\begin{array}{ccc}
M^{(2)}\left( s_2,\mu ,\nu \right) & = & \frac 1{2\,\pi \left| \nu \right|
}\left| \int_{\omega >0} 
\frac{d\omega }{\sqrt{\omega }}f(\omega )e^{-i\left( \frac \mu \nu \omega -%
\frac{s_2}\nu \log \omega \right) }\right| ^2+ \\  &  & \frac 1{2\,\pi
\left| \nu \right| }\left| \int_{\omega <0}\frac{d\omega }{\sqrt{\left|
\omega \right| }}f(\omega )e^{-i\left( \frac \mu \nu \omega -\frac{s_2}\nu
\log \left| \omega \right| \right) }\right| ^2 
\end{array}
\end{equation}
$f(\omega )$ being the Fourier transform of the signal $f(t)$.

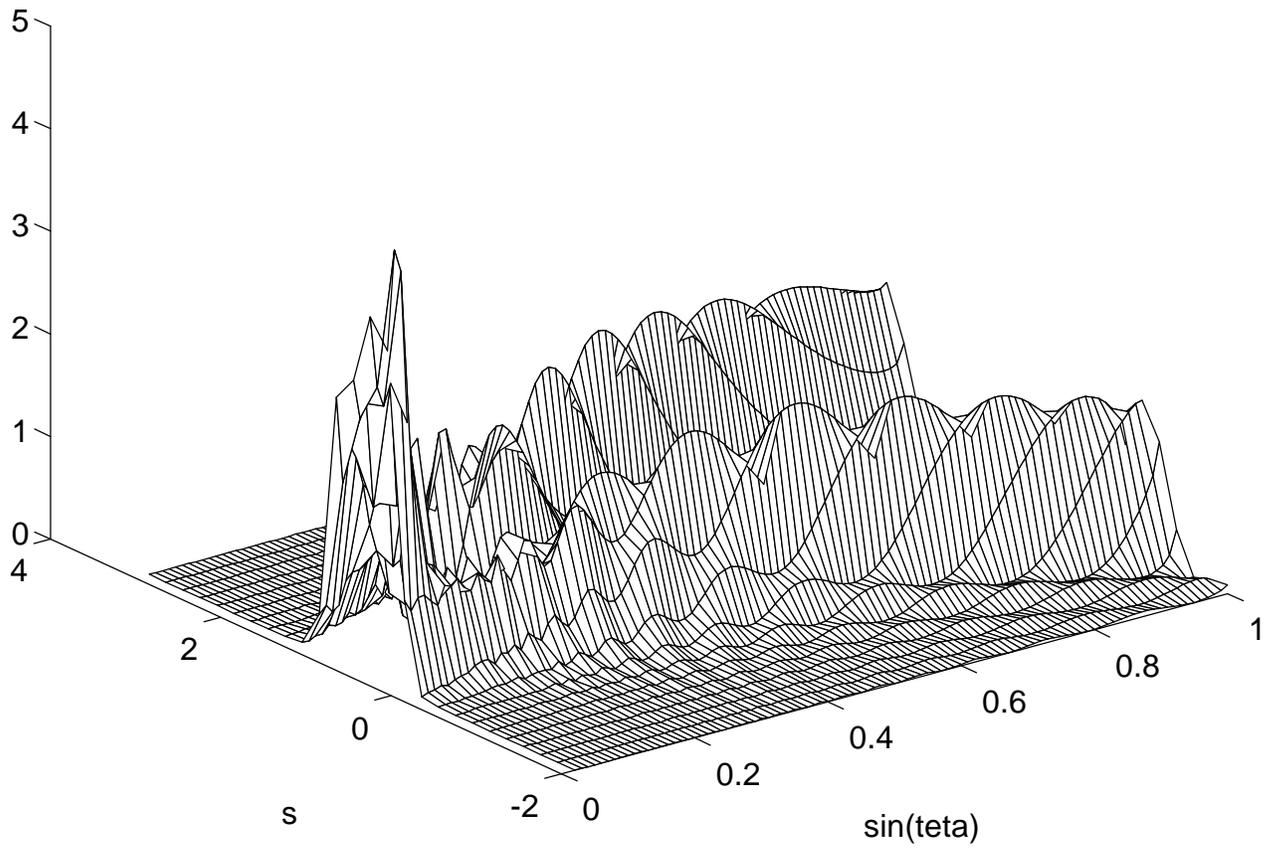
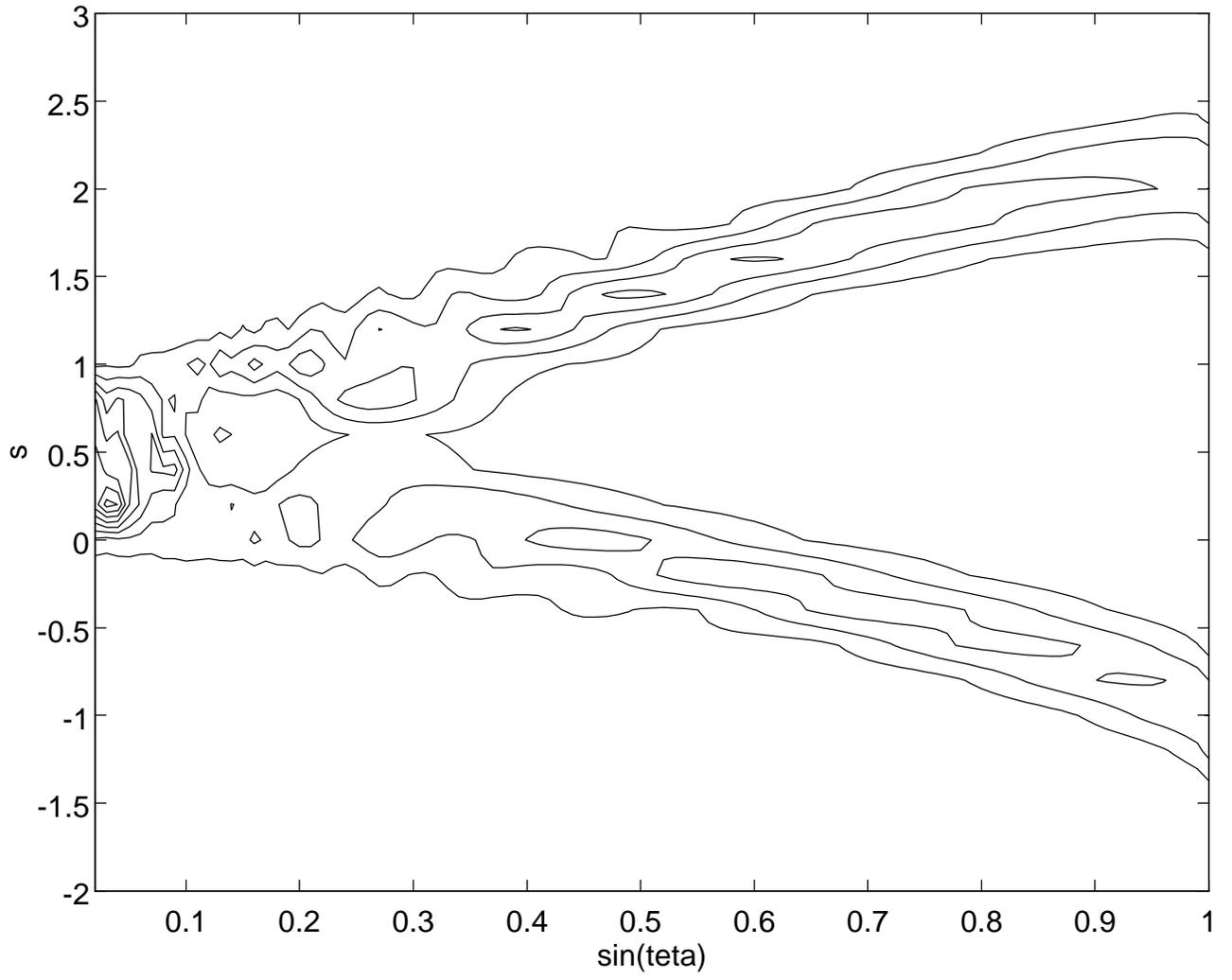
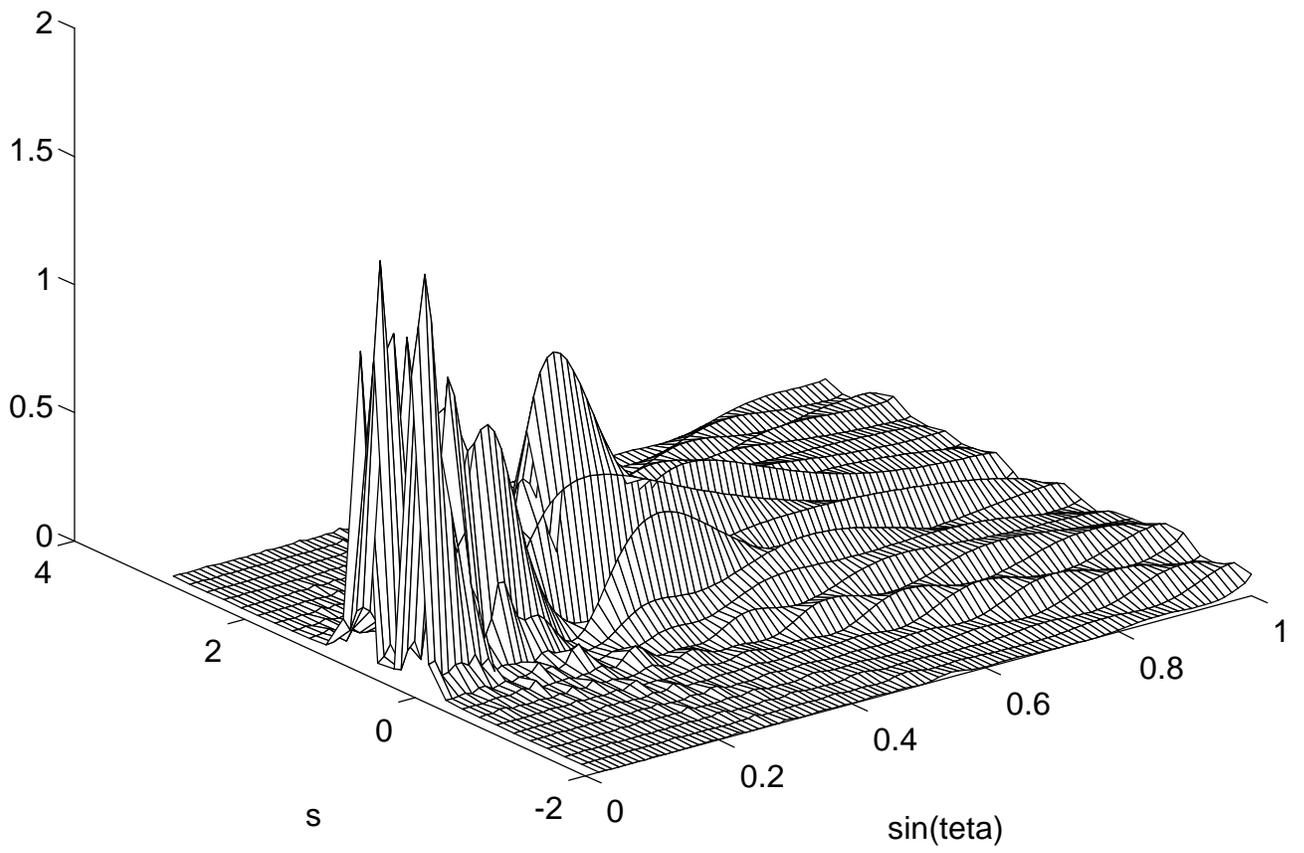
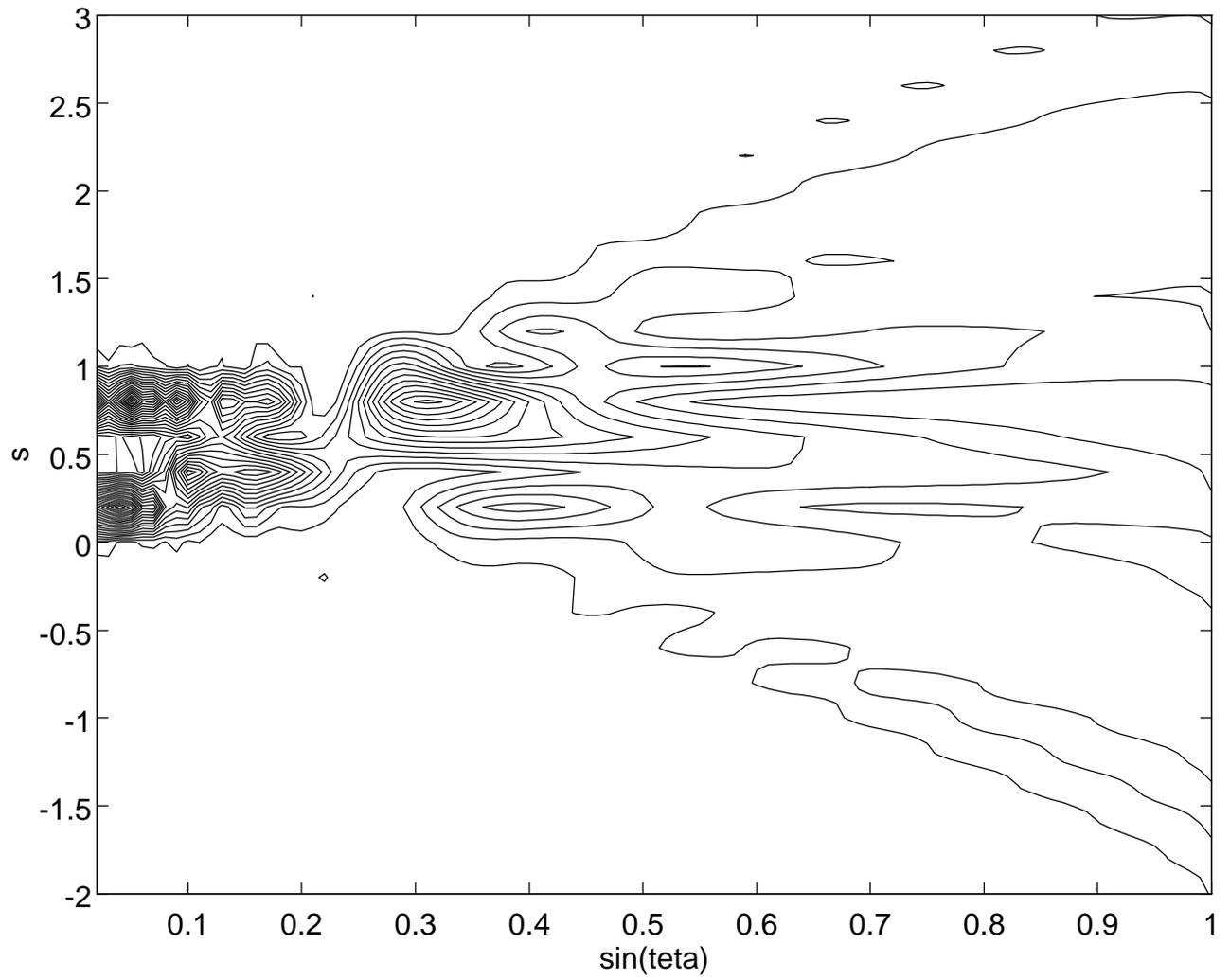
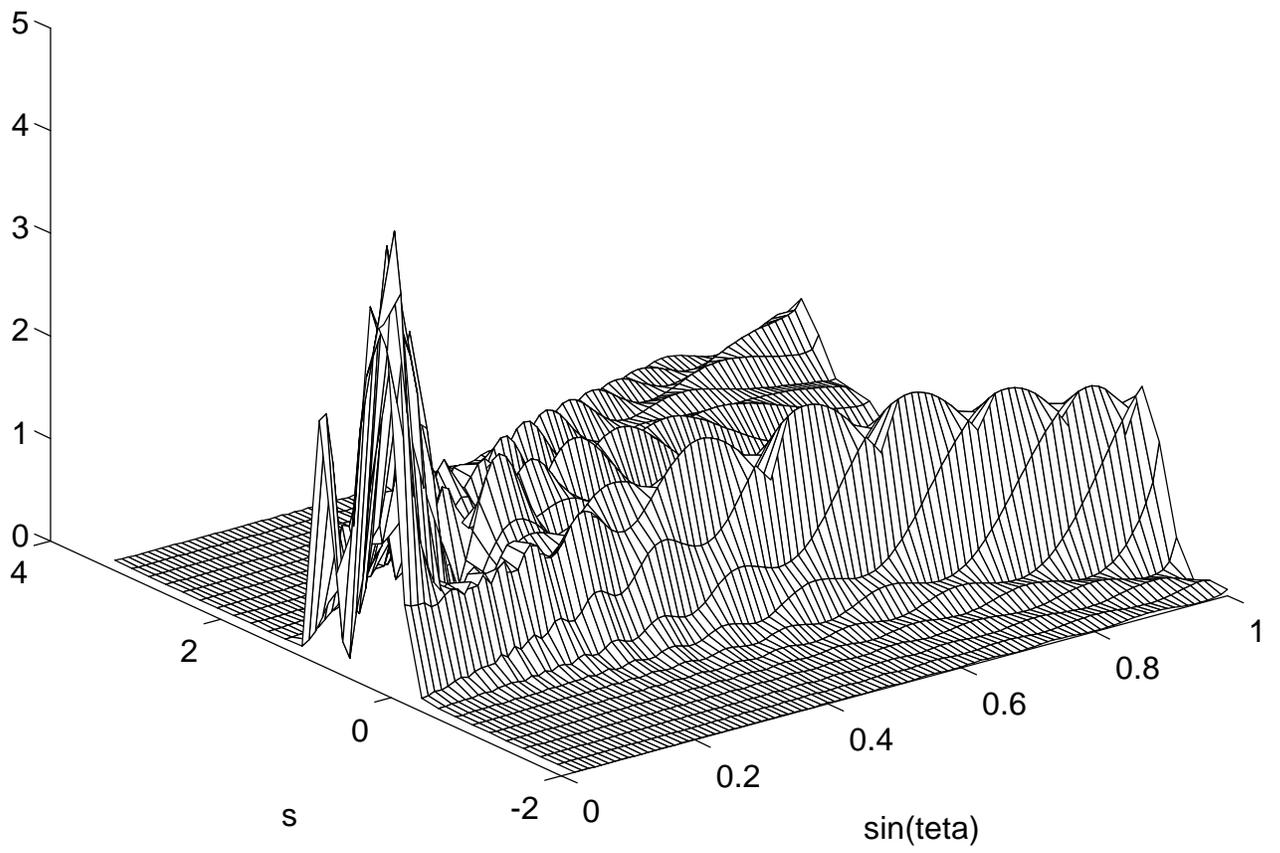
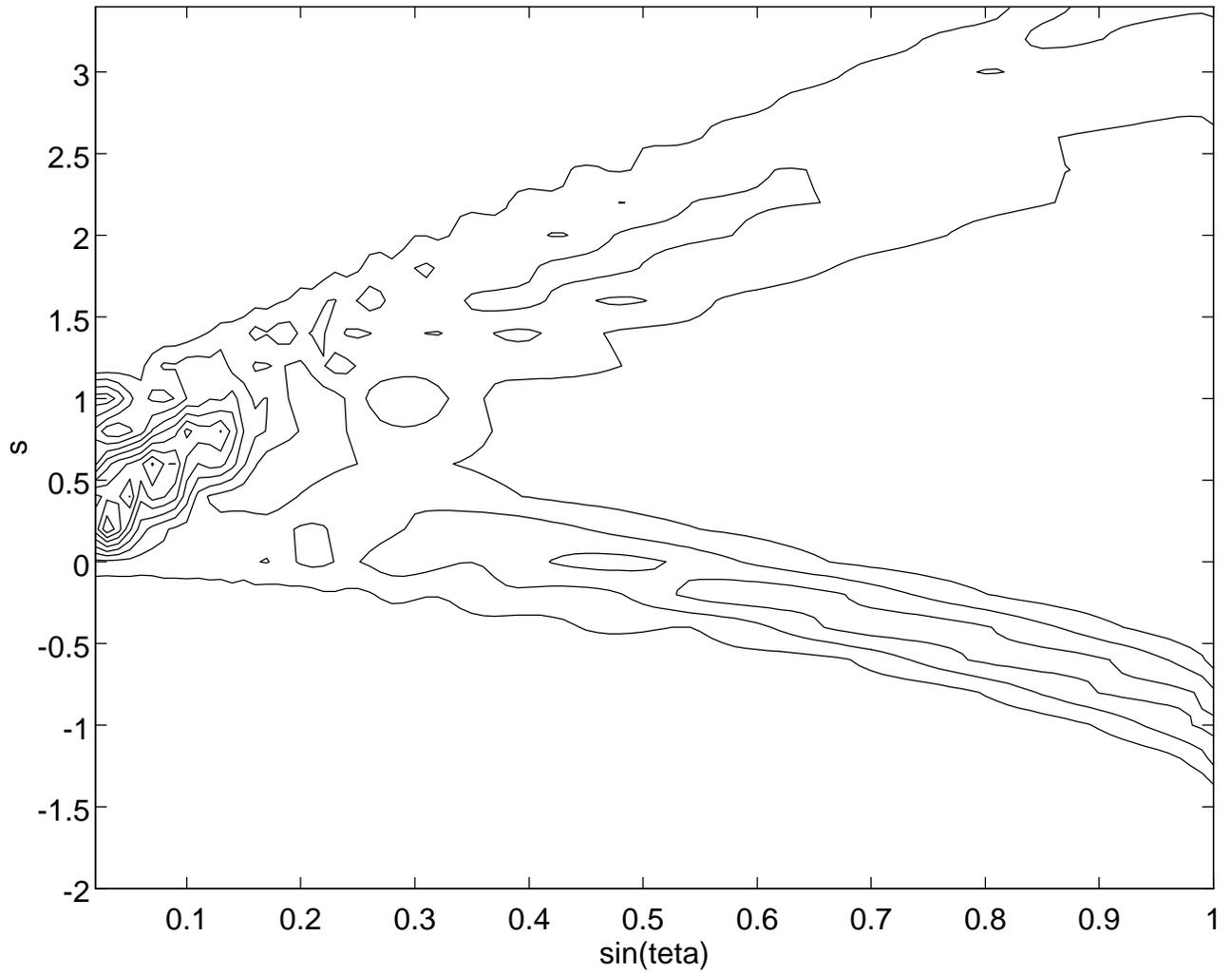
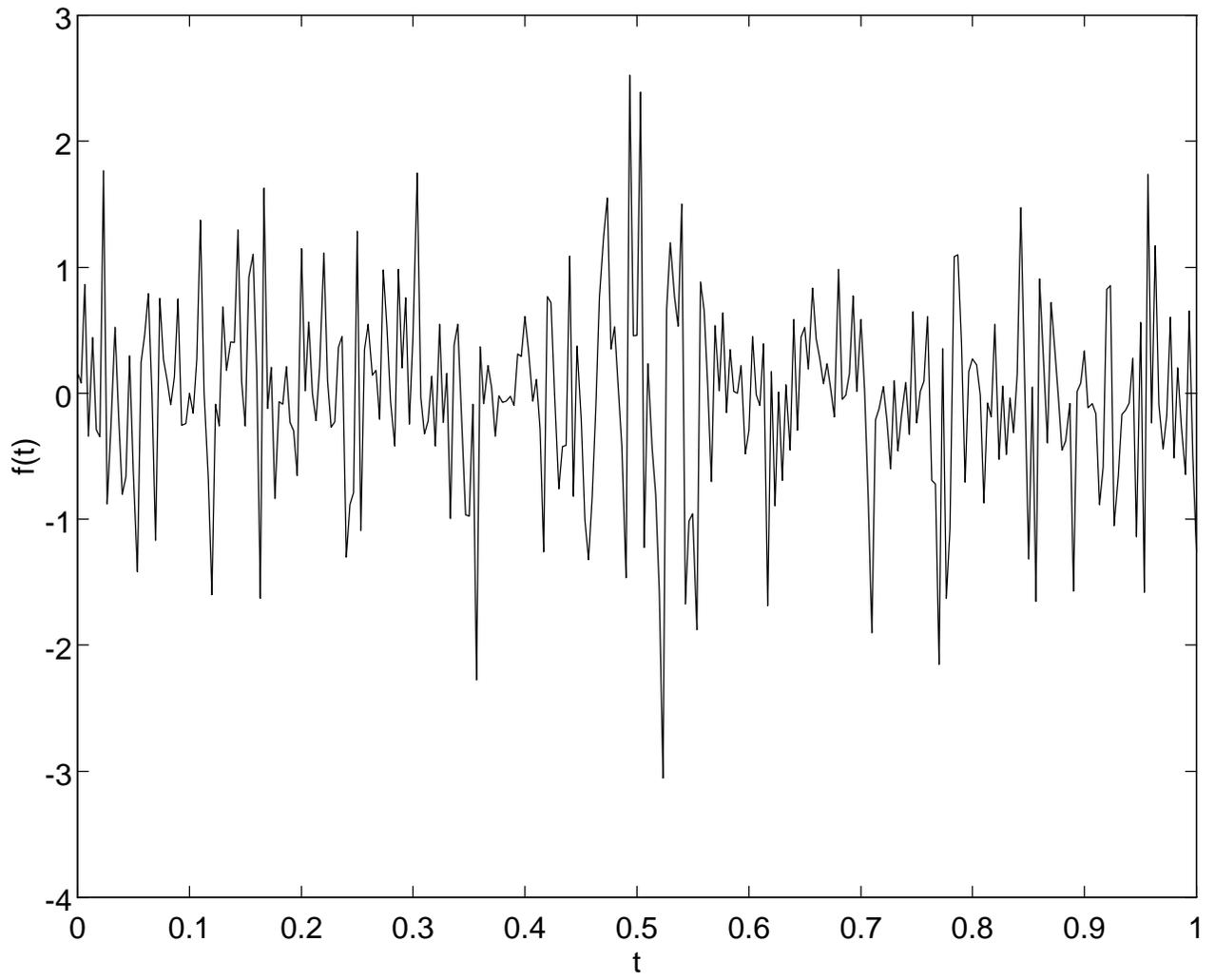
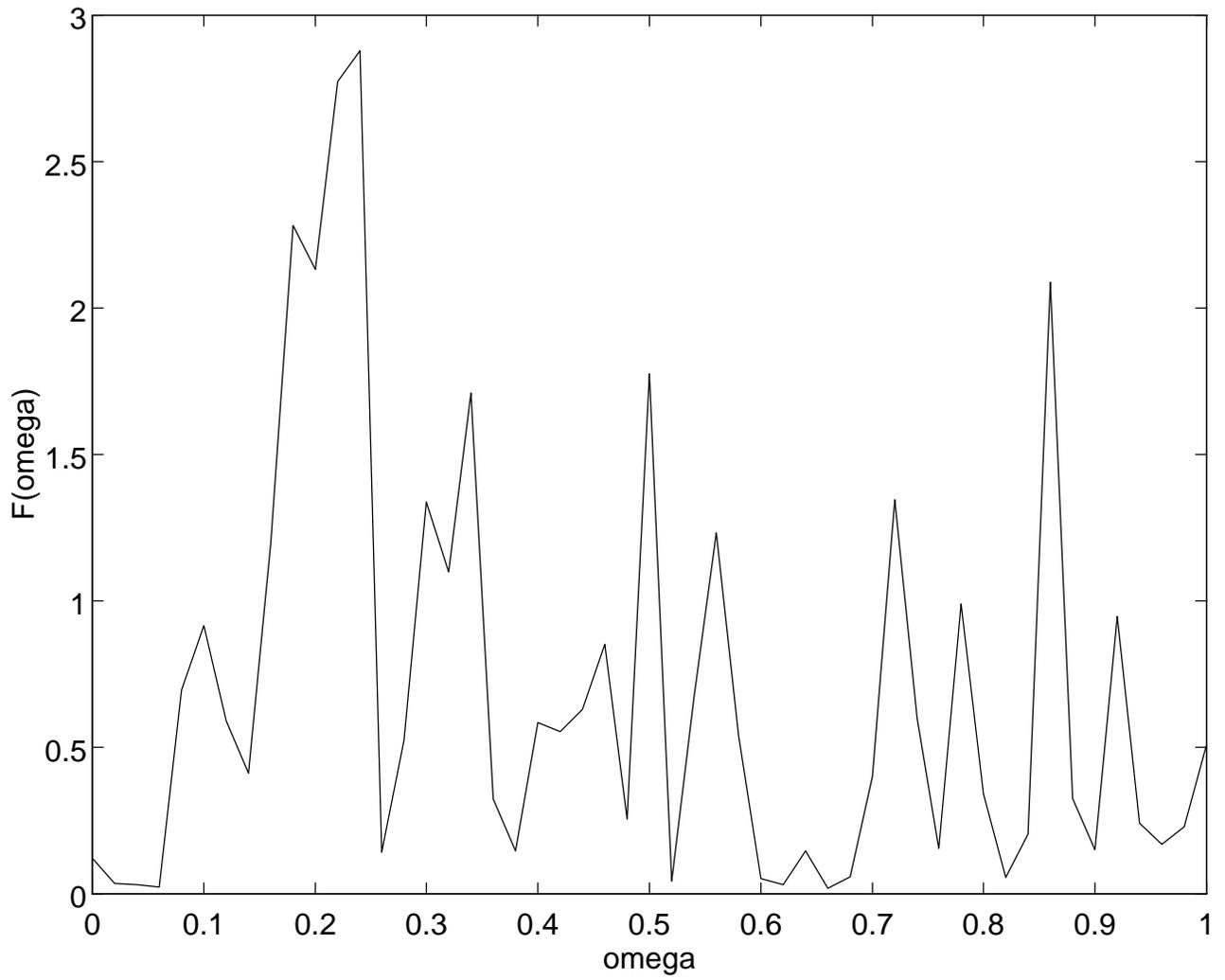
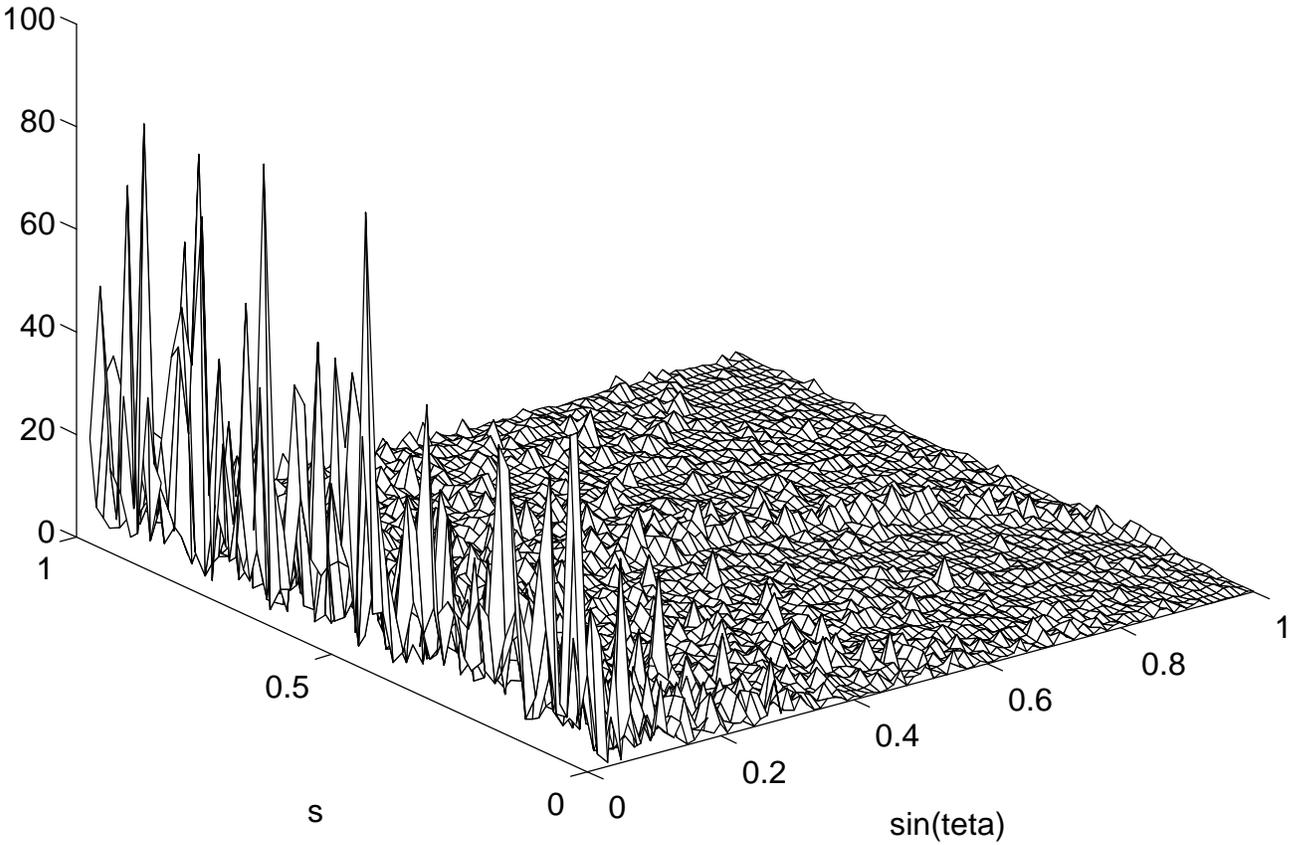
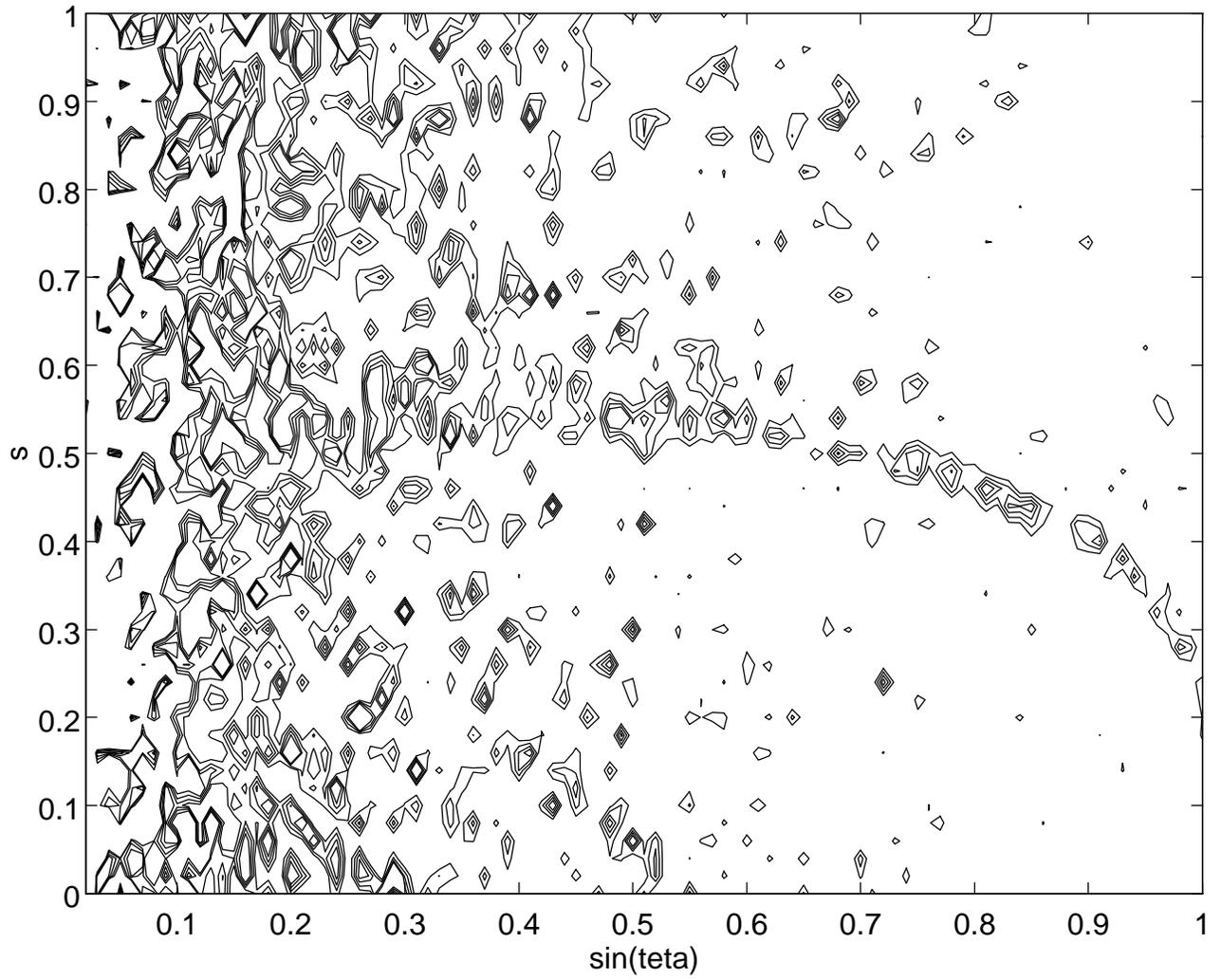
\end{document}